\documentclass[10pt]{article}
\usepackage{graphicx}
\usepackage{xspace}
\usepackage{amsmath}
\usepackage{lineno}

\def\Title#1{\begin{center} {\Large #1 } \end{center}}
\def\Author#1{\begin{center}{ \sc #1} \end{center}}
\def\Address#1{\begin{center}{ \it #1} \end{center}}

\newcommand\pubblock{\rightline{\begin{tabular}{l} Proceedings of the Second Annual LHCP\\ \pubnumber\\
         \pubdate  \end{tabular}}}

\newenvironment{Abstract}{\begin{quotation} \begin{center} 
             \large ABSTRACT \end{center}\bigskip 
      \begin{center}\begin{large}}{\end{large}\end{center} \end{quotation}}

\newenvironment{Presented}{\begin{quotation} \begin{center} 
             PRESENTED AT\end{center}\bigskip 
      \begin{center}\begin{large}}{\end{large}\end{center} \end{quotation}}

\def\beq{\begin{equation}}
\def\eeq#1{\label{#1}\end{equation}}
\def\eeqn{\end{equation}}

\def\beqa{\begin{eqnarray}}
\def\eeqa#1{\label{#1}\end{eqnarray}}
\def\eeqan{\end{eqnarray}}

\let\bar=\overbar

\def\D{{\cal D}}

\def\Dslash{\not{\hbox{\kern-4pt $D$}}}
\def\dslash{\not{\hbox{\kern-2pt $\del$}}}

\def\ee{e^+e^-}

\def\msb{{\bar{\ssstyle M \kern -1pt S}}}

\def\neu#1{\widetilde\chi^0_{#1}}

\def\lhcb {LHCb\xspace}
\newcommand{\decay}[2]{\ensuremath{{#1\!\to #2}}\xspace}
\newcommand{\tev}{\ensuremath{\mathrm{\,Te\kern -0.1em V}}\xspace}
\newcommand{\gev}{\ensuremath{\mathrm{\,Ge\kern -0.1em V}}\xspace}
\newcommand{\mev}{\ensuremath{\mathrm{\,Me\kern -0.1em V}}\xspace}
\newcommand{\kev}{\ensuremath{\mathrm{\,ke\kern -0.1em V}}\xspace}
\newcommand{\ev}{\ensuremath{\mathrm{\,e\kern -0.1em V}}\xspace}
\newcommand{\gevc}{\ensuremath{{\mathrm{\,Ge\kern -0.1em V\!/}c}}\xspace}
\newcommand{\mevc}{\ensuremath{{\mathrm{\,Me\kern -0.1em V\!/}c}}\xspace}
\newcommand{\gevcc}{\ensuremath{{\mathrm{\,Ge\kern -0.1em V\!/}c^2}}\xspace}
\newcommand{\gevgevcccc}{\ensuremath{{\mathrm{\,Ge\kern -0.1em V^2\!/}c^4}}\xspace}
\newcommand{\mevcc}{\ensuremath{{\mathrm{\,Me\kern -0.1em V\!/}c^2}}\xspace}
\def\invfb   {\ensuremath{\mbox{\,fb}^{-1}}\xspace}
 \def\PJ      {\ensuremath{\mathrm{J}}\xspace}                 
 \def\Ppsi        {\ensuremath{\psi}\xspace}                 
\def\jpsi     {\ensuremath{{\PJ\mskip -3mu/\mskip -2mu\Ppsi\mskip 2mu}}\xspace}
 \def\Ps      {\ensuremath{\mathrm{s}}\xspace}                 
\def\squark    {\ensuremath{\Ps}\xspace}
\def\squarkbar {\ensuremath{\overline \squark}\xspace}
 \def\Pmu         {\ensuremath{\mu}\xspace}                 
 \def\Ppi         {\ensuremath{\pi}\xspace}                 
\def\pion  {\ensuremath{\Ppi}\xspace}
\def\Kp    {\ensuremath{\kaon^+}\xspace}
\def\Km    {\ensuremath{\kaon^-}\xspace}

\def\mup        {\ensuremath{\Pmu^+}\xspace}
\def\mun        {\ensuremath{\Pmu^-}\xspace} 
\def\pip   {\ensuremath{\pion^+}\xspace}
\def\pim   {\ensuremath{\pion^-}\xspace}
\newcommand{\BsPP}{\decay{\Bs}{\phi \phi}}

\newcommand{\BsJPPP}{\decay{\Bs}{\jpsi \pi^+ \pi^-}}

\def\CP                {\ensuremath{C\!P}\xspace}

 \def\PB      {\ensuremath{B}\xspace}                 
                  
 \def\Pc      {\ensuremath{c}\xspace}                 
                  
\def\B       {\ensuremath{\PB}\xspace}
  \def\Bbar    {\kern 0.18em\overline{\kern -0.18em \PB}{}\xspace}

\def\Bu      {\ensuremath{\B^+}\xspace}

\def\Bp      {\ensuremath{\Bu}\xspace}

\def\Bd      {\ensuremath{\B^0}\xspace}
\def\Bs      {\ensuremath{\B^0_\squark}\xspace}
\def\Bsb     {\ensuremath{\Bbar^0_\squark}\xspace}
\newcommand{\DGs}{\ensuremath{\Delta\Gamma_{\squark}}\xspace}
\newcommand{\Gs}{\ensuremath{\Gamma_{\squark}}\xspace}
\newcommand{\GL}{\ensuremath{\Gamma_{\rm L}}\xspace}
 \def\PK      {\ensuremath{\mathrm{K}}\xspace}                 
 \def\Pb      {\ensuremath{\mathrm{b}}\xspace}                 
\def\ssbar     {\ensuremath{\squark\squarkbar}\xspace}
\def\bquark    {\ensuremath{\Pb}\xspace}
\def\bquarkbar {\ensuremath{\overline \bquark}\xspace}
\def\cquark    {\ensuremath{\Pc}\xspace}
\def\cquarkbar {\ensuremath{\overline \cquark}\xspace}
\def\ccbar     {\ensuremath{\cquark\cquarkbar}\xspace}
\newcommand{\GH}{\ensuremath{\Gamma_{\rm H}}\xspace}
\def\invps{\ensuremath{{\rm \,ps^{-1}}}\xspace}
\newcommand{\dms}{\ensuremath{\Delta m_{\squark}}\xspace}
\def\rad{\ensuremath{\rm \,rad}\xspace}
\newcommand{\phisC}{\ensuremath{\phi_{\rm s}^{\ccbar \squark}\xspace}\xspace}
\newcommand{\phisS}{\ensuremath{\phi_{\rm s}^{\ssbar \squark}\xspace}\xspace}
\def\kaon  {\ensuremath{\PK}\xspace}
\def\Kstarz  {\ensuremath{\kaon^{*0}}\xspace}
\newcommand{\stat}{\ensuremath{\mathrm{(stat)}}\xspace}
\newcommand{\syst}{\ensuremath{\mathrm{(syst)}}\xspace}
\def\ps   {\ensuremath{{\rm \,ps}}\xspace}

\def\be{\begin{equation}}
\def\ee{\end{equation}}
\def\bea{\begin{eqnarray}}
\def\eea{\end{eqnarray}}

\def\Lb      {\ensuremath{\Lambda_\bquark}\xspace}
 \def\Pp      {\ensuremath{p}\xspace}                 
\def\proton      {\ensuremath{\Pp}\xspace}
\newcommand{\etaprime}{\ensuremath{\eta^{\prime}}\xspace}
\def\Kstarz  {{\ensuremath{\kaon^{*0}}}\xspace}
\def\D       {{\ensuremath{D}}\xspace}
\def\Dz      {{\ensuremath{\D^0}}\xspace}
\def\neu        {{\ensuremath{\nu}}\xspace}
\def\neum       {{\ensuremath{\neu_\mu}}\xspace}
\def\Dstm    {{\ensuremath{\D^{*-}_\squark}}\xspace}
\def\Dsm     {{\ensuremath{\D^-_\squark}}\xspace}
\def\pipm   {{\ensuremath{\pion^\pm}}\xspace}

\usepackage{color}

\newcommand\pubnumber{ LHCB-PROC-2014-036 }

\newcommand\pubdate{\today}

\def\affiliation{
CERN, Geneva, Switzerland.}

\def\support{\footnote{ on behalf of the \lhcb collaboration }}

\begin{document}
\large
\begin{titlepage}
\pubblock

\vfill
\Title{  \CP violation in \Bs decays at \lhcb  }
\vfill

\Author{ Sean Benson \support }
\Address{\affiliation}
\vfill
\begin{Abstract}
Latest \lhcb measurements of \CP violation in the interference between mixing and decay are presented based
on $pp$ collision data collected during LHC Run~I, corresponding to an 
integrated luminosity of 3.0\invfb.
Approximately $27\, 000$ $\Bs\to \jpsi \pip\pim$ signal events are used to make what is at the moment the most
precise single measurement of the \CP-violating phase in $\bquark\to\ccbar \squark$ transitions, $\phisC=0.070\pm0.068\stat\pm0.008\syst$\rad.
The most accurate measurement of the \CP-violating phase in $\bquark\to \ssbar\squark$ transitions, \phisS, is
found from approximately $4\, 000$ $\Bs\to \phi\phi$ signal events 
to be $\phisS=-0.17\pm0.15\stat\pm0.03\syst$\rad.
\end{Abstract}
\vfill

\begin{Presented}
The Second Annual Conference\\
 on Large Hadron Collider Physics \\
Columbia University, New York, U.S.A \\ 
June 2-7, 2014
\end{Presented}
\vfill
\end{titlepage}
\def\thefootnote{\fnsymbol{footnote}}
\setcounter{footnote}{0}
%

\normalsize 


\section{Introduction}
\label{sec:Introduction}

Efforts to measure mixing-induced \CP violation in the \Bs system have mainly focused on 
the $\Bs \to \jpsi \phi$ decay, utilising angular observables to disentangle
the \CP-odd and \CP-even components. This then allows for the \CP-violating phase, \phisC,
to be measured. In the Standard Model, $\phisC \approx -2\beta_s
= 2arg(- V_{ts} V_{tb}^* / V_{cs} V_{cb}^*)$~\cite{Faller:2008gt,Dighe:1998vk,Dunietz:2000cr}. The Standard Model
prediction for \phisC has been obtained from global fits to experimental data yielding 
a value of $-0.036\pm 0.002$~\rad~\cite{Charles:2011va,Lenz:2006hd,Lenz:2011ti}. There are however many New Physics theories 
that provide additional contributions to \Bs mixing diagrams which alter this value~\cite{Ball:2006xx,Lenz:2007nj}. 
The addition of the \BsJPPP decay allows for an independent
determination of \phisC. 

The \CP-violating phase measured in the $\Bs \to \phi \phi$ decay results from 
$\bquark \to \ssbar \squark$ transitions and is therefore expected to be close to zero 
in the Standard Model due to the effective cancellation of the \CP-violating weak phase
between the \Bs mixing diagrams and the penguin decay diagrams~\cite{Raidal:2002ph,Bhattacharya:2013sga}.
Calculations using QCD factorisation provide an upper limit of 0.02\rad for $|\phisS|$~\cite{Bartsch:2008ps,Cheng:2009mu}.

The following sections summarise updated measurements of the \CP-violating weak phases in $\bquark \to \ccbar \squark$ and
$\bquark \to \ssbar \squark$ transitions from the full LHCb Run~I dataset 
of 3.0\invfb, using $\Bs\to (\jpsi\to \mup\mun) \pip\pim$ and $\Bs \to (\phi \to \Kp\Km ) \; (\phi \to \Kp\Km )$ decays, respectively~\cite{Aaij:2014dka,Aaij:2014kxa}.

\section{The $\mathbf{\Bs \to \jpsi \pi \pi}$ analysis}
\label{sec:jpsipipi}

Previous analyses measuring \CP violation in $\bquark\to\ccbar\squark$ transitions have been made using
\lhcb data collected in 2011, consisting of 1.0\invfb, where the combined measurement
of the \CP-violating phase, \phisC, was found to be $0.01\pm0.07\stat\pm0.01\syst\rad$~\cite{Aaij:2013oba,LHCb:2012ad}.
While previous analyses have used the measured result that the two-pion invariant mass spectrum
is almost entirely \CP-odd~\cite{LHCb:2012ae}, the updated result presented here~\cite{Aaij:2014dka}, uses $27\,100\pm200$ signal 
events and incorporates an amplitude analysis that avoids assumptions on the \CP content.

\begin{figure}[h]
\centering
\includegraphics[width=0.35\textwidth]{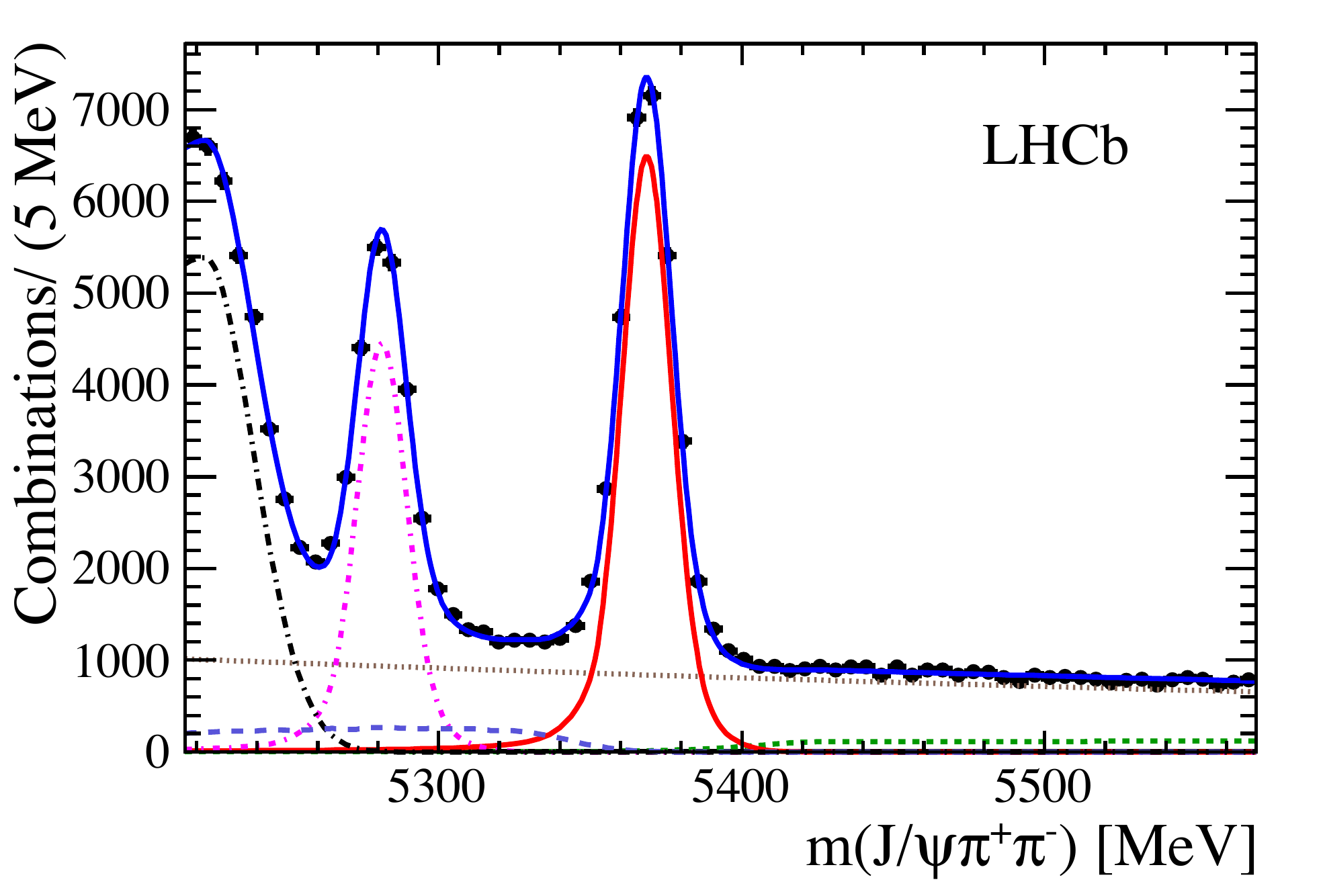}
  \caption{\small Distribution of the $\jpsi\pip\pim$ invariant mass. Data are represented by black markers,
  the dotted magenta and solid red lines denote the $\Bd\to \jpsi\pip\pim$ and $\Bs\to \jpsi\pip\pim$ fit components, respectively.
  The dotted brown line and the blue solid line represent the combinatorial background and the total fit, respectively.
  The reflections from $\Bd\to\jpsi\Kp\pim$ and $\Bp\to\jpsi h^{+}$ decays are given by the dotted black and green lines, respectively,
  and the dashed blue line represents the sum of the $\Bs \to \jpsi \etaprime$, $\Bs\to\jpsi (\phi\to \pip\pim\pi^{0})$, and $\Lb\to\jpsi\Km\proton$ reflections.
  }
  \label{fig:mass_pipi}
\end{figure}

Figure~\ref{fig:mass_pipi} shows the four-particle invariant mass range, $m(\jpsi\pip\pim)$, from which the
shape of the combinatorial background component is determined. For the \CP violation measurement, events 
in the range $|m(\jpsi\pip\pim)-m_{\Bs}|<20\mevcc$ are taken, such that only the \Bs signal and combinatorial
background components are present, where $m_{\Bs}$ is the PDG \Bs mass. The $\Bs\to\jpsi\pip\pim$ component
is modelled with a double Crystal Ball function~\cite{Skwarnicki:1986xj}, with the combinatorial background modelled with an
exponential.

In order to measure \CP violation an un-binned maximum-likelihood fit is performed to the invariant mass, 
$m(\jpsi\pip\pim)$, the $\pip\pim$ invariant mass, $m(\pip\pim)$, the three helicity angles defined in Ref.~\cite{Aaij:2014dka}, 
$\Omega\in \{\theta_{\pi\pi},\theta_{\jpsi},\chi\}$ and the decay time, $t$. 
The total decay rates for the $\Bs\to\jpsi\pip\pim$ and $\Bsb\to\jpsi\pip\pim$ decays, denoted by  $\Gamma(t)$ and 
$\overline{\Gamma}(t)$ respectively, can be written as
\begin{align}
\Gamma(t) &= e^{-\Gs t}\left( \frac{|\mathcal{A}|^2+|\overline{\mathcal{A}}|^2}{2} \cosh\frac{\Gs t}{2} +
\frac{|\mathcal{A}|^2-|\overline{\mathcal{A}}|^2}{2}\cos(\dms t) \right. \nonumber\\
&\left. -\Re(\mathcal{A}^*\overline{\mathcal{A}})\sinh\frac{\Gs t}{2} - \Im(\mathcal{A}^*\overline{\mathcal{A}})\sin(\dms t)\right), \label{eq:pipi_G1}\\
\overline{\Gamma}(t) &= |p/q|^2e^{-\Gs t}\left( \frac{|\mathcal{A}|^2+|\overline{\mathcal{A}}|^2}{2} \cosh\frac{\Gs t}{2} - \frac{|\mathcal{A}|^2-|\overline{\mathcal{A}}|^2}{2}\cos(\dms t) \right. \nonumber\\
&\left. -\Re(\mathcal{A}^*\overline{\mathcal{A}})\sinh\frac{\Gs t}{2} + \Im(\mathcal{A}^*\overline{\mathcal{A}})\sin(\dms t)\right), \label{eq:pipi_G2}
\end{align}
where $\Gs\equiv(\GL+\GH)/2$ is the average decay rate, $\DGs\equiv\GL-\GH$ is the decay rate difference and \dms is 
the oscillation frequency of the \Bs system. The decay amplitudes are defined as $\mathcal{A} = \sum_i A_i$ and
$\overline{\mathcal{A}}=\sum_i \lambda_i A_i$, where $\lambda_i \equiv (q/p)(\overline{A}_i/A_i)$, $A_i$  and 
$\overline{A}_i$ are the amplitudes of \Bs mesons and \Bsb mesons to the final state, $i$, at $t=0$, and the 
complex parameters, $q$ and $p$, relate the flavour eigenstates to the mass eigenstates of the \Bs system
at $t=0$.
The full helicity dependence of the amplitudes on the two-pion invariant mass and helicity angles is provided in
 Ref.~\cite{Zhang:2012zk}.
The probability density function (PDF) includes detector resolution and acceptance effects. The complete PDF is factorised
to separate the $\jpsi\pip\pim$ invariant mass from the other observables. The values of \Gs, \DGs and \dms are
constrained to \lhcb measurements~\cite{Aaij:2013oba,Aaij:2013mpa}. 

\begin{figure}[h]
\begin{center}
\includegraphics[width=0.35\textwidth]{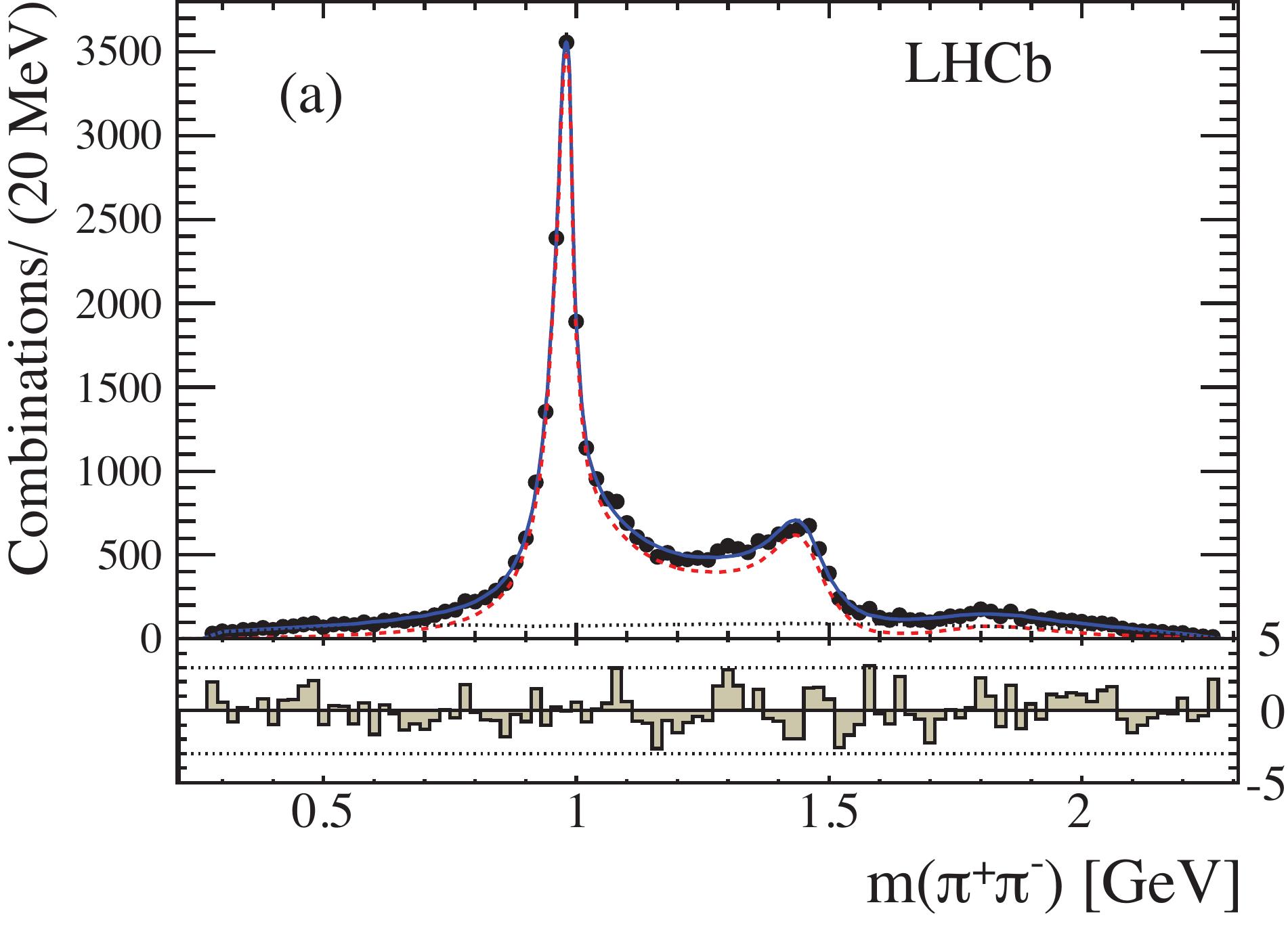}
\includegraphics[width=0.35\textwidth]{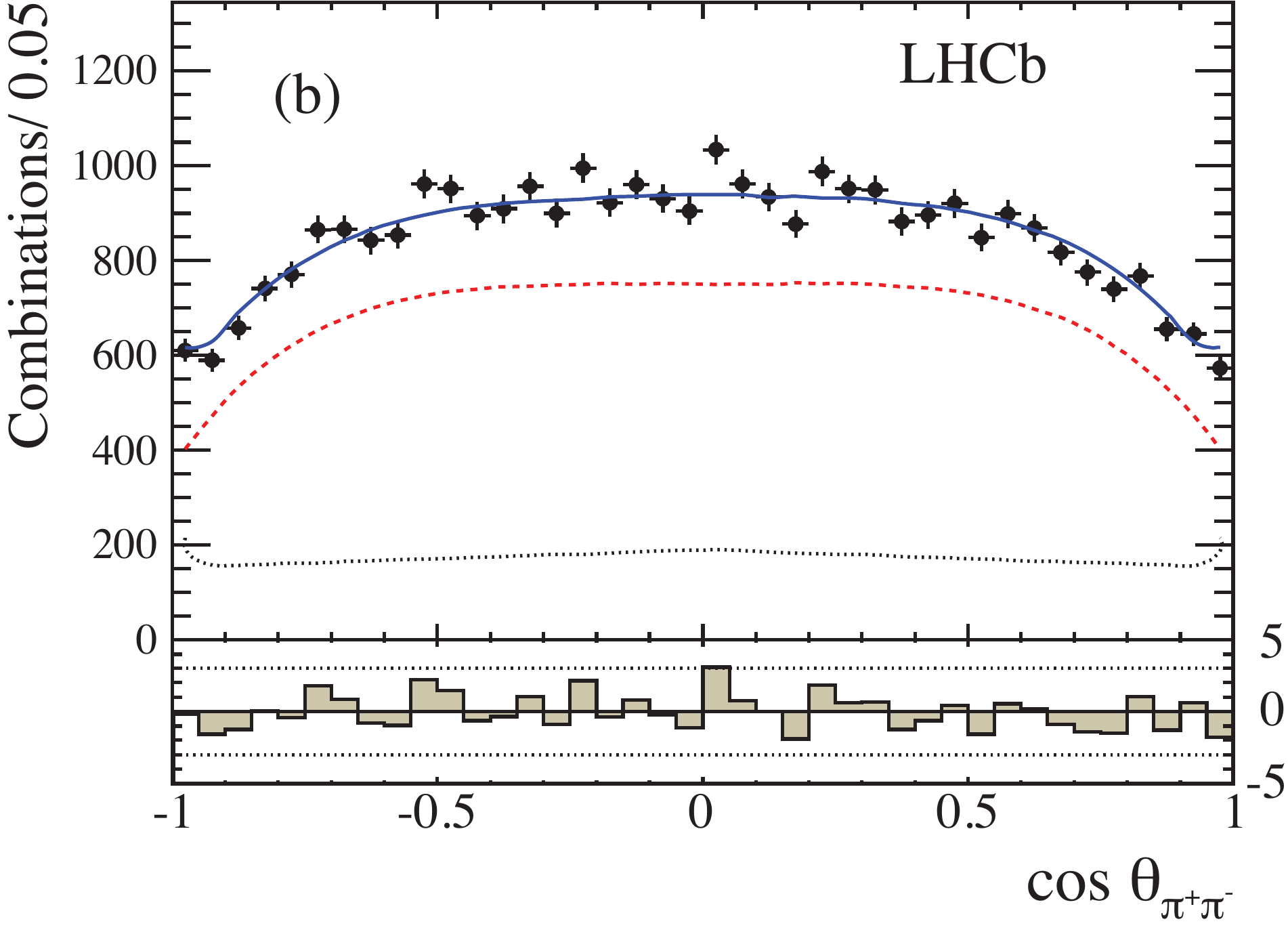}\\
\includegraphics[width=0.35\textwidth]{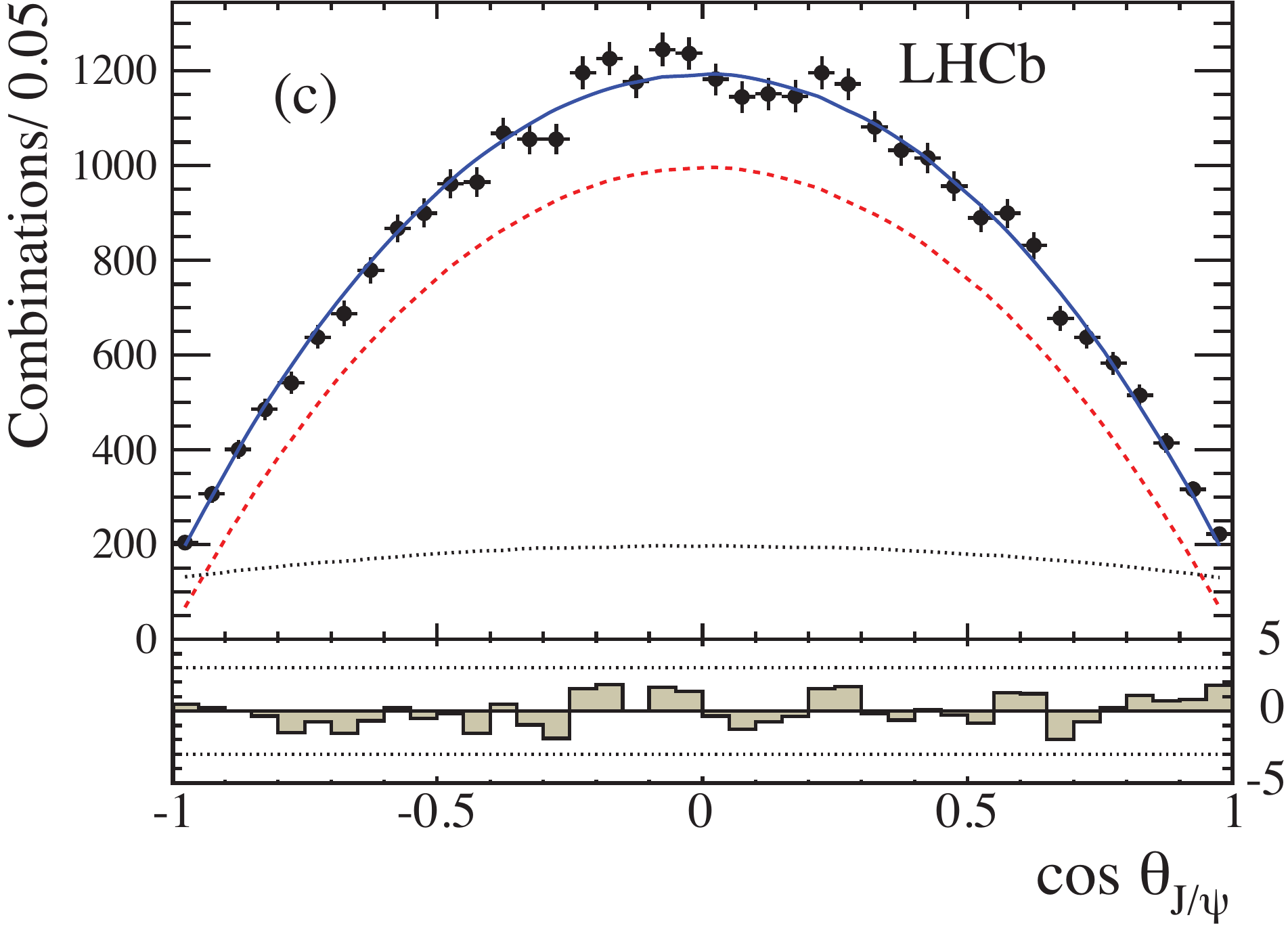}
\includegraphics[width=0.35\textwidth]{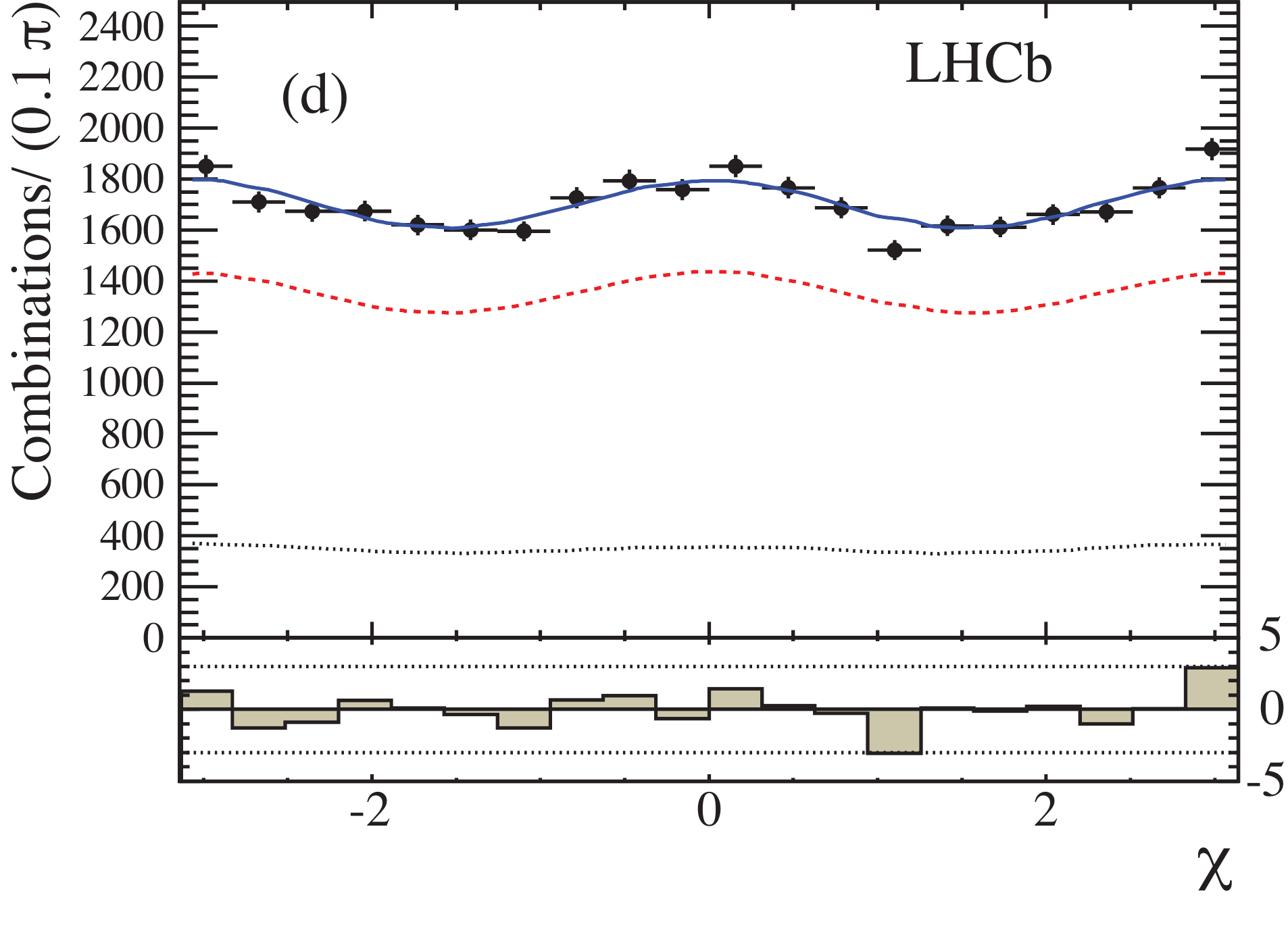}
  \caption{
  \small Projections of (a) $m(\pip\pim)$, (b) $\cos\theta_{\pi\pi}$, (c) $\cos\theta_{\jpsi}$, and (d) $\chi$. Data
  are shown by black markers, the total fit, signal and combinatorial background components are given by the solid blue line,
  dashed red line, and dotted black lines respectively.
  }
  \label{fig:angles_pipi}
\end{center}
\end{figure}

It can be seen from Eqs.~\ref{eq:pipi_G1} and \ref{eq:pipi_G2} that knowledge of the initial flavour of
the \Bs meson at production provides extra sensitivity to \CP violation. 
At \lhcb, so-called flavour tagging is achieved through the use of the algorithms described in Refs.~\cite{Aaij:2013oba,SSKConf}. 
This analysis uses both the opposite side (OS) and same side kaon (SSK) flavour taggers. The OS flavour tagging algorithm~\cite{FTConf} makes use of the $\bquarkbar(\bquark)$-quark produced in association 
with the signal $\bquark(\bquarkbar)$-quark. The predicted probability of an incorrect flavour assignment, $\omega$, 
is determined for each event by a neural network that is calibrated using $\Bu\to\jpsi\Kp$, $\Bu\to\Dz\pip$, $\Bd\to\jpsi\Kstarz$, $\Bd\to \Dstm\mup\neum$, and $\Bs\to\Dsm\pip$ data as flavour
specific control modes. 
Details of the calibration procedure can be found in Ref.~\cite{Aaij:2013oba}. When a signal \Bs meson is formed, 
an associated \squark-quark is produced in the fragmentation that 
forms a charged kaon around 50\,\% of the time, The aforementioned charged kaon is likely to originate close to the \Bs meson production point. 
The kaon charge therefore allows for the identification of the flavour of the signal \Bs meson. 
This principle is exploited by the SSK flavour tagging algorithm~\cite{SSKConf}. 
The overall tagging power, calculated as $\epsilon_{\rm tag}(1-2\omega_{\rm avg})^2$, is found to be $(3.89\pm0.25)\,\%$,
where $\epsilon_{\rm tag}$ is the tagging efficiency, and $\omega_{\rm avg}$ is the average wrong-tag probability.

Figure~\ref{fig:angles_pipi} shows the projections of the two-pion invariant mass and the helicity angles. Good fit quality is 
seen showing that the complex $m(\pip\pim)$ spectrum comprising of the $f_0(980)$, $f_0(1500)$, $f_0(1790)$, 
$f_2(1270)$, and $f_2^\prime(1525)$ resonances and associated interferences is understood. Efficiencies as a 
function of $m(\pip\pim)$ and $\Omega$ are obtained from simulated events. The background distributions of the 
helicity angles are taken as the sum of the individual contributions and are parameterised as described in Ref.~\cite{Aaij:2014emv}.

\begin{figure}[h]
\begin{center}
\includegraphics[width=0.35\textwidth]{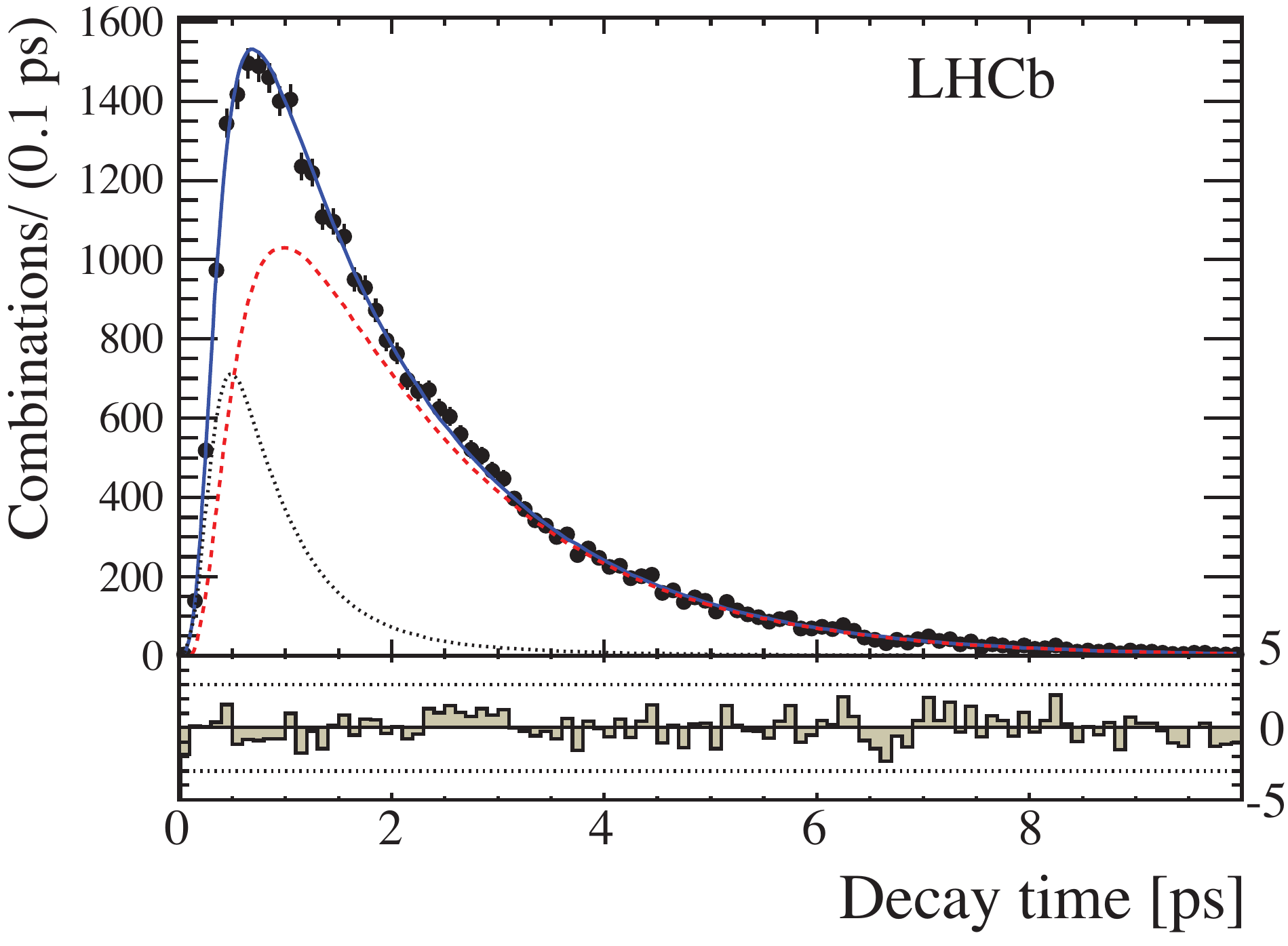}
  \caption{
    \small Decay time distribution of $\Bs\to\jpsi \pip\pim$ candidates. Data
      are shown by black markers, the total fit, signal and combinatorial background components are given by the solid blue line,
        dashed red line, and dotted black lines respectively.
    }
  \label{fig:time_pipi}
\end{center}
\end{figure}

Figure~\ref{fig:time_pipi} shows the projection of the \Bs decay time integrated over all other observables, for events
inside a 40\mevcc window centred on the PDG \Bs mass. The decay time 
acceptance is obtained from $\Bd\to\jpsi\Kstarz$ events. The decay time resolution makes use of the per-event decay time
error which is obtained from the kinematics of the candidate in question and is used in a triple-Gaussian model, after 
calibration using prompt $\jpsi \to \mup\mun$ candidates.  
The background distribution is described using like-sign $\pipm\pipm$ candidates to obtain the parameters
of a double exponential distribution combined with an acceptance function of the form 
$[a(t-t_0)]^n/(1+[a(t-t_0)]^n)\times(1+\beta t + \beta_2 t^2)$, where $a,\, t_0,\, n,\, \beta$ and $\beta_2$ are parameters
to be fitted.

The dominant contributions to the systematic uncertainty are found to be from the production asymmetry and the
models used to parameterise the resonances in the $m(\pip\pim)$ spectrum. The former is accounted for by
multiplying $\overline{\Gamma}(t)$ by the \Bs/\Bsb production ratio, $R_p=(1.00\pm0.05)$~\cite{Norrbin:2000jy}
and varying $R_p$ within the associated error. The uncertainty on the resonance model arises from the addition of a $\rho(770)$ 
component, even though this component is forbidden by isospin conservation. The uncertainties on the \CP-violating
phase from these sources are 0.006\rad in both cases. The uncertainties on the direct \CP violation
parameter from these sources are 0.002 and 0.010, respectively.

The result of the measurement of the weak phase $\phi_s$ in the \BsJPPP decay is found to
be $\phi_s=0.070\pm0.068\stat\pm0.008$\rad~\cite{Aaij:2014dka}. The direct \CP violation parameter, $|\lambda|$, is measured to be 
$0.89\pm0.05\stat\pm0.01\syst$~\cite{Aaij:2014dka} (note that a value of unity signifies no direct \CP violation). This result is more precise than the previous 
measurement using 1.0\invfb of \lhcb data and is the most accurate single measurement of the \CP-violating
phase in $\bquark \to \ccbar \squark$ transitions.
%

\section{The $\mathbf{\BsPP}$ Analysis}
\label{sec:phiphi}

\begin{figure}[t]
\begin{center}
\includegraphics[width=0.35\textwidth]{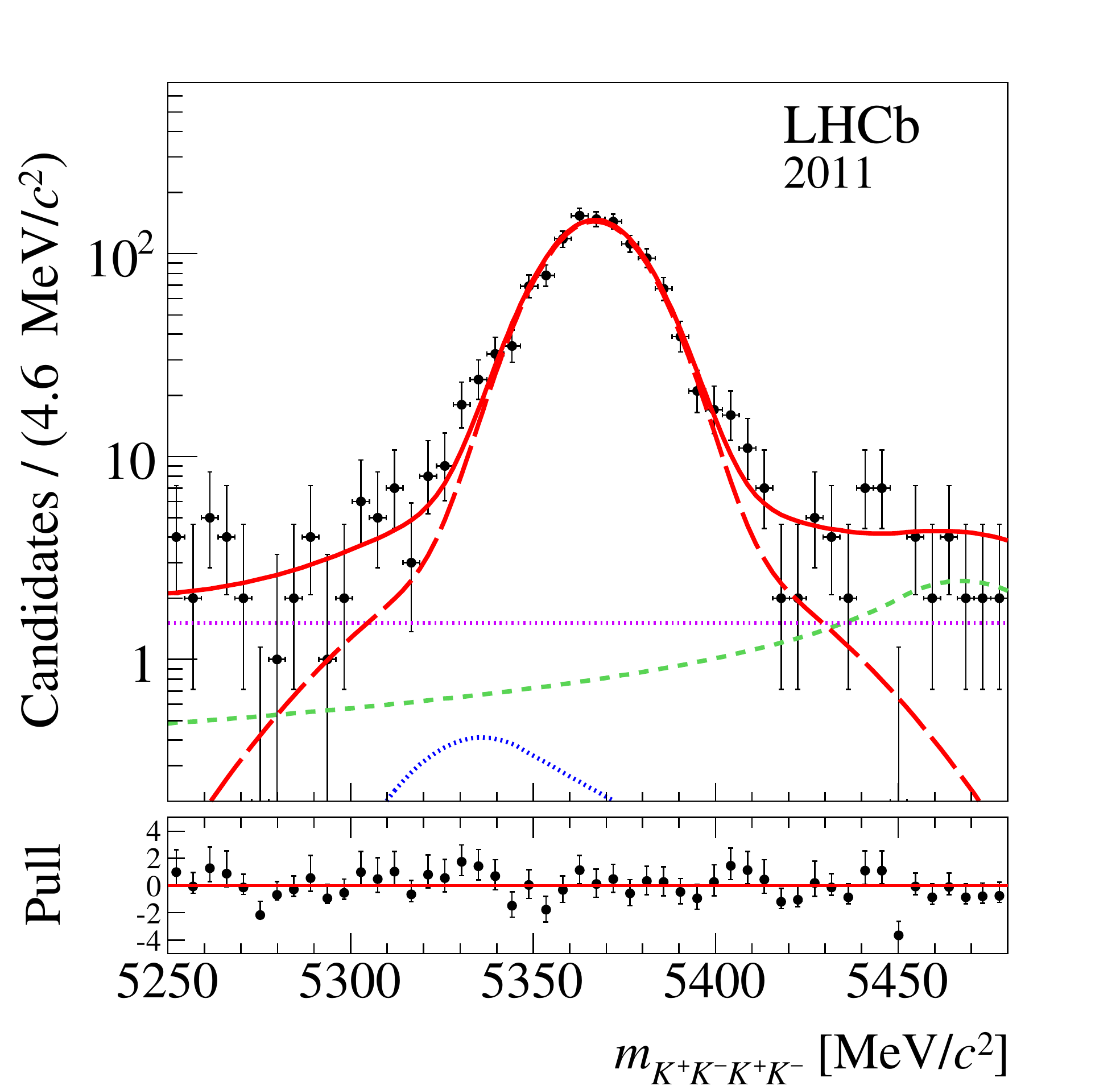}
\includegraphics[width=0.35\textwidth]{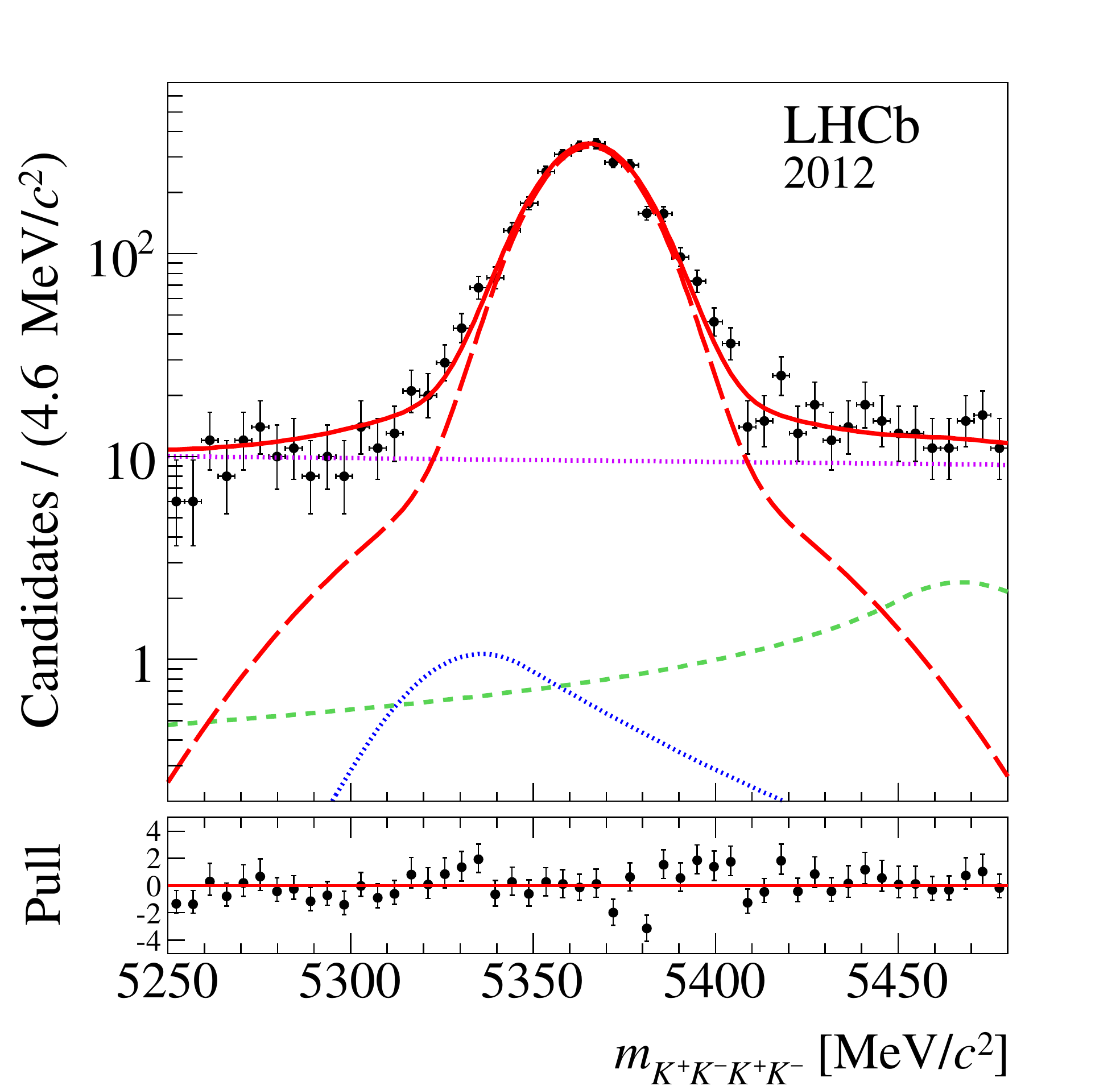}
  \caption{
  \small Four-kaon invariant mass distributions for the (left) 2011 and (right) 2012 datasets.
  Data are represented by black markers. Superimposed are the results of the total fit
  (red solid line), the \BsPP (red long dashed), the $\Bd\to\phi\Kstarz$ (blue dotted),
  the $\Lb\to\phi \proton\Km$ (green short-dashed), and the combinatoric (purple dotted) fit components.}
  \label{fig:mass_phiphi}
\end{center}
\end{figure}

The \BsPP decay is an example of a flavour changing
neutral current (FCNC) decay and as such, may only proceed via penguin diagrams in the Standard Model.
The most recent analysis builds on the previous first measurement of the \CP-violating phase~\cite{Aaij:2013qha},
in addition to the measurement of the triple product asymmetries~\cite{Aaij:2012ud} using 1.0\invfb of \lhcb data.
A total of $4000$ signal candidates are observed through a multivariate selection to distinguish signal from background. Figure~\ref{fig:mass_phiphi} shows the $\Kp\Km\Kp\Km$ invariant mass after all selections have been applied.

As in the case of the \BsJPPP analysis, a maximum log-likelihood fit is then performed to the three helicity angles 
and to the decay time.
The \BsPP decay is a  $P\to VV$ decay, where $P$ denotes a pseudoscalar and $V$ a vector meson.
However, due to the proximity of the $\phi$ 
resonance to that of the $f_0(980)$, 
there will also be contributions from $S$-wave ($P\to V\kern-0.1em S$) and double $S$-wave ($P\to S\kern-0.1em S$) processes, where $S$ denotes 
a spin-0 meson or a pair of non-resonant kaons.
Thus the total amplitude is a coherent sum of $P$-, $S$-, and double $S$-wave processes, and is
accounted for during fitting
by making use of the different functions of the helicity angles associated with these terms. The functional form 
of the PDF in terms of the decay time and helicity angles is given in Ref.~\cite{Aaij:2014kxa}.
The parameters of interest are the \CP violation parameters ($\phisS$ and $|\lambda|$), the polarisation amplitudes ($|A_0|^2$, $|A_\perp|^2$, $|A_S|^2$, and $|A_{SS}|^2$), 
and the \CP-conserving strong phases ($\delta_1$, $\delta_2$, $\delta_S$, and $\delta_{SS}$). 
The $P$-wave amplitudes are defined such that
$|A_0|^2+|A_\perp|^2+|A_\parallel|^2=1$, hence only two are free parameters.

Flavour tagging is achieved with the same algorithms as used for the measurement of \phisC in the \BsJPPP decay.
The efficiencies as a function of decay angles are accounted for with simulated \BsPP events that have been
subjected to the same selection requirements as used for the data sample. The decay time acceptance is accounted for
with the $\Bs\to\Dsm\pip$ control mode, that is re-weighted according to the final state particle transverse momentum.
The same decay time biasing selections as used to select the \BsPP decay are applied in addition to a requirement that the 
\Dsm decay time be less than $1\ps$,
to enforce topological similarity to the \BsPP decay. Decay time resolution is accounted for with a per-event decay time 
error, used in association with a Gaussian model after having first been calibrated using simulated events.

The 2011 and 2012 data samples are assigned independent signal weights, decay time and angular acceptances,
in addition to separate Gaussian constraints to the decay time resolution parameters.
The value of the \Bs-\Bsb oscillation frequency is constrained to the \lhcb measured value
~\cite{Aaij:2013mpa}. The values of the decay
width and decay width difference are constrained to the \lhcb measured values
of $\Gs=0.661\pm0.004\stat\pm0.006\syst$\invps and $\DGs=0.106\pm0.011\stat\pm0.007\syst$\invps, 
respectively~\cite{Aaij:2013oba}.
The Gaussian constraints applied to the \Gs and \DGs parameters use the combination of the measured
values from $\Bs\to\jpsi\Kp\Km$ and $\Bs\to\jpsi\pip\pim$ decays. Constraints are therefore applied
taking into account a correlation of $0.1$ for the statistical uncertainties~\cite{Aaij:2013oba}.
The systematic uncertainties are taken to be uncorrelated between the $\Bs\to\jpsi\Kp\Km$ and $\Bs\to\jpsi\pip\pim$ decay modes.

\begin{figure}[t]
\begin{center}
\includegraphics[width=0.35\textwidth]{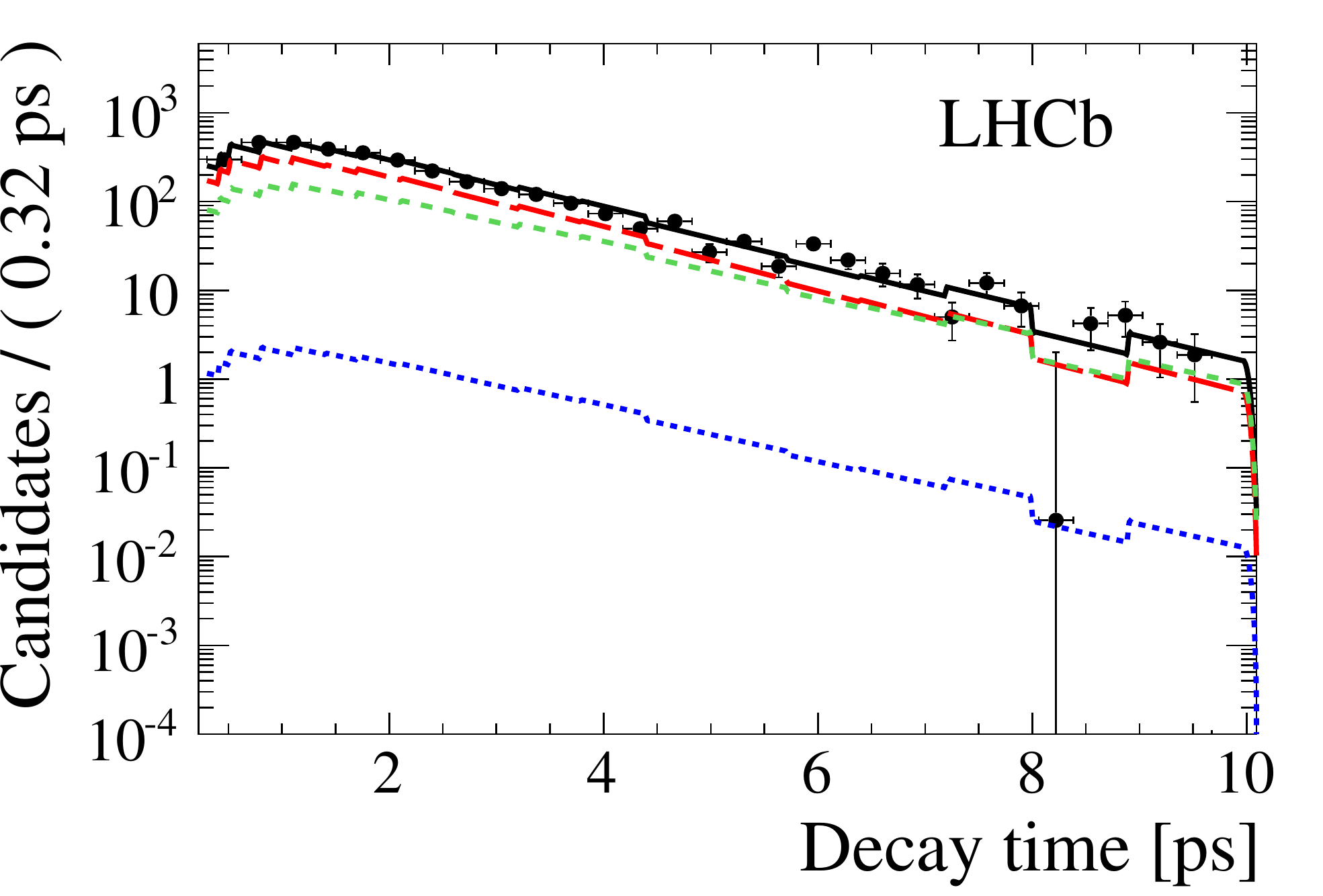}
\includegraphics[width=0.35\textwidth]{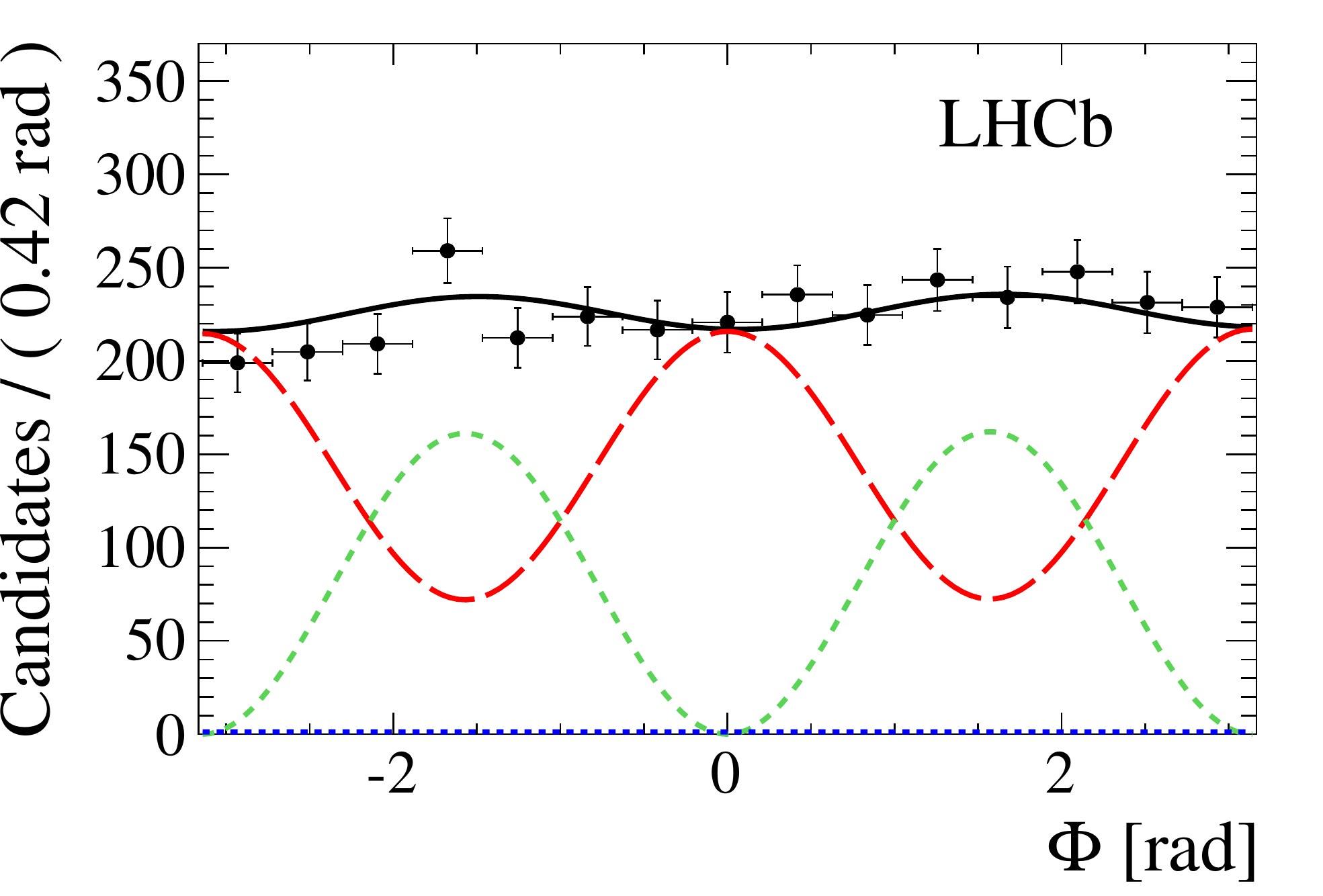}\\
\includegraphics[width=0.35\textwidth]{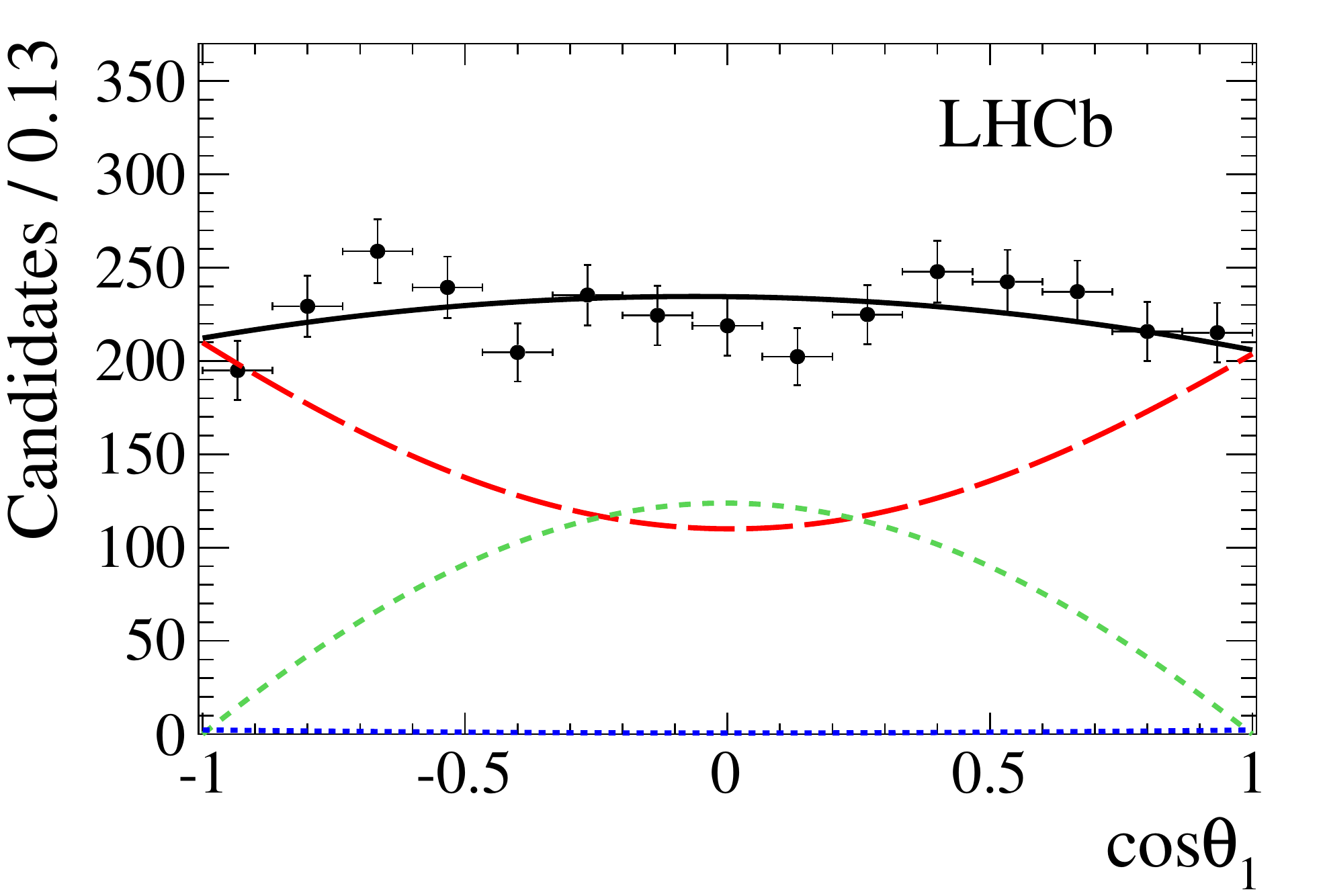}
\includegraphics[width=0.35\textwidth]{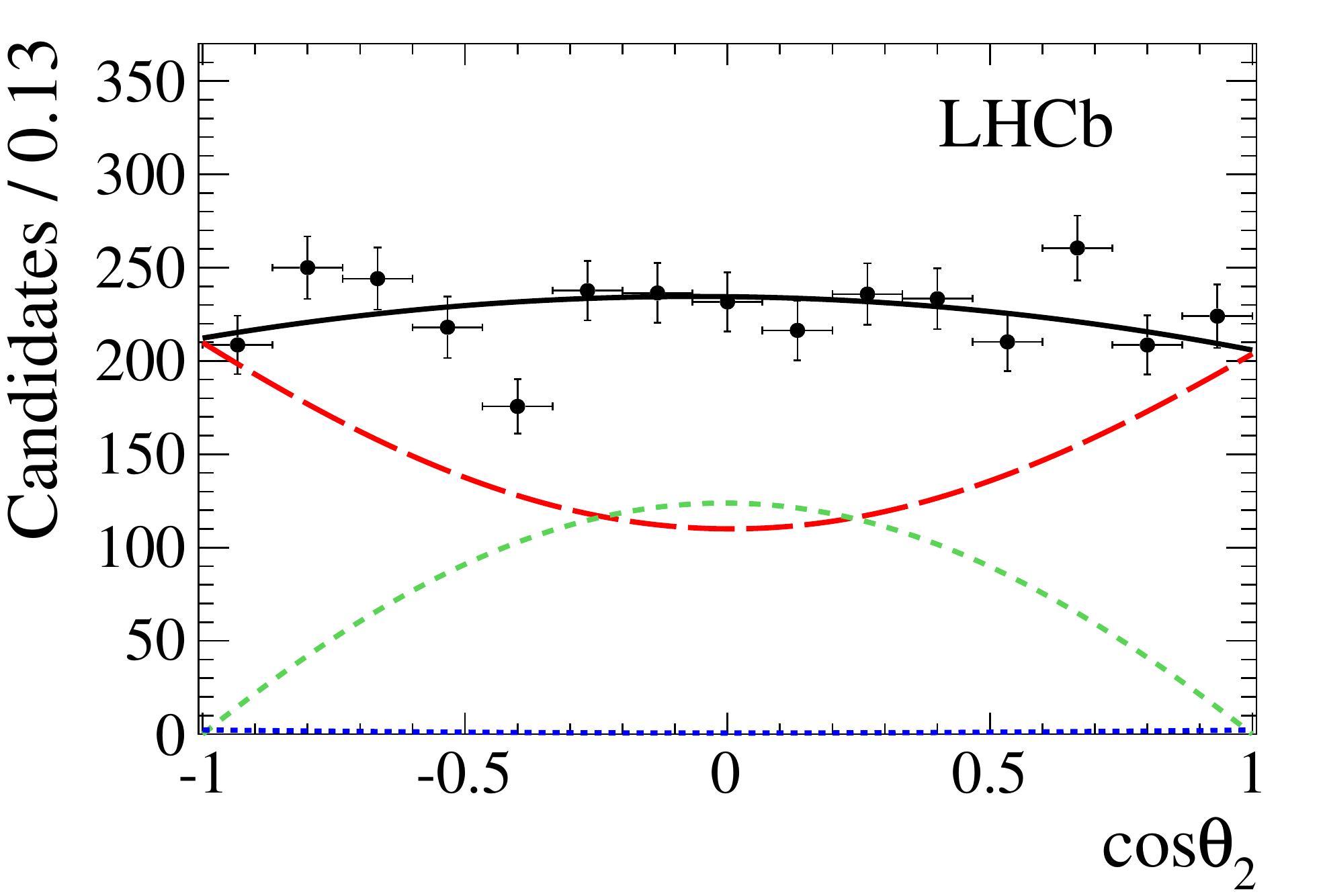}
  \caption{
  One-dimensional projections of the $\Bs \to \phi\phi$ fit for (top-left) decay time with binned acceptance,
    (top-right) helicity angle $\Phi$ and (bottom-left and bottom-right) cosine
        of the helicity angles $\theta_1$ and $\theta_2$.
               The background-subtracted data are marked as black points, while the black solid lines represent the projections of the best fit.
                         The \CP-even $P$-wave, the \CP-odd $P$-wave and $S$-wave combined with double $S$-wave
                                components are shown by the red long dashed, green short dashed and blue dotted lines, respectively.}
  \label{fig:angles_phiphi}
\end{center}
\end{figure}
Figure~\ref{fig:angles_phiphi} shows the projections on to the helicity angles and decay time,
the fit to which yields \CP violation parameters of $\phisS=-0.17\pm0.15\stat\pm0.03\syst$\rad and
$|\lambda|=1.04\pm0.07\stat\pm0.03\syst$. Polarisation amplitudes are measured to be
$|A_0|^2=0.364\pm0.012\stat\pm0.009\syst$ and $|A_\perp|^2=0.305\pm0.013\stat\pm0.005\syst$,
where $|A_\parallel |^2$ is fixed such that the fractions of the $P$-wave sum to unity. In addition,
the $S$-wave fractions are found to be consistent with a pure $P$-wave state.

A separate un-binned maximum log-likelihood fit is performed to the four-kaon mass in data samples that have been divided
according to the sign of the $T$-odd observables, $U=\sin\Phi\cos\Phi$ and $V=\pm\sin\Phi$, where the positive
sign is taken if $\cos\theta_1\cos\theta_2>0$ else the negative sign is used~\cite{Gronau:2011cf}.
With such a fit, asymmetries can be measured in the $T$-odd observables, denoted $A_U$ and $A_V$, which
provide a method of measuring \CP violation that does not require flavour tagging or knowledge
of the decay time.
These so-called triple product asymmetries are measured to be $A_U=-0.003\pm0.017\stat\pm0.006\syst$ and $A_V=-0.017\pm0.017\stat\pm0.006\syst$.  

The dominant sources of systematic uncertainties are found to arise from the angular and decay time acceptances,
which each contribute uncertainties of 0.02\rad to the systematic uncertainty on \phisS.
The mass model is also found to have a significant effect for the measurement of the triple product asymmetries. 

\section{Summary}
\label{sec:summary}

The most accurate single measurement of \CP violation in \Bs mixing has been presented
using the full Run~I dataset collected with the \lhcb detector, corresponding to an
integrated luminosity of 3.0\invfb.
The analysis of approximately $27,\,000$ \BsJPPP decays yields a measurement
of $\phisC=0.070\pm0.068\stat\pm0.008\syst$\rad. 
The most precise measurement of the \CP-violating phase in the 
\BsPP penguin decay is also presented, which is found to be $\phisS=-0.17\pm0.15\stat\pm0.03\syst$\rad.
All results are consistent with Standard Model expectations. Statistical uncertainties are found
to be dominant in all measurements of \CP violation, hence measurements with greater precision can be 
expected with the addition of Run~II data. 

\bibliographystyle{srddfdft}

\end{document}